\documentclass{aa}

\usepackage{natbib}
\bibpunct{(}{)}{;}{a}{}{,}
\usepackage{graphicx}
\usepackage{txfonts}

\begin{document}
\title{Band profiles and band strengths in mixed H$_2$O:CO ices}
\titlerunning{Band profiles and band strengths in mixed H$_2$O:CO ices}

\newcommand{\water}{H$_2$O }
\newcommand{\co}{CO$_2$ }
\newcommand{\wav}{cm$^{-1}$ }

\author{Jordy Bouwman \inst{1}
          \and Wiebke Ludwig \inst{1}
          \and Zainab Awad \inst{1,3} 
          \and Karin I. \"Oberg  \inst{1}
          \and Guido W. Fuchs \inst{1}
          \and Ewine F. van Dishoeck  \inst{2}
          \and Harold Linnartz  \inst{1}
          }

   \offprints{Jordy Bouwman, bouwman@strw.leidenuniv.nl}

  \institute{Raymond and Beverly Sackler Laboratory for Astrophysics, Leiden Observatory, Leiden University, P.O. Box 9513, NL 2300 RA Leiden, The Netherlands
\and Leiden Observatory, Leiden University, P.O. Box 9513, NL 2300 RA Leiden, The Netherlands \and Visiting scientist from Department of Astronomy, Faculty of Science, Cairo University, Egypt}

   \date{Received ; accepted}


  \abstract
   {Laboratory
   spectroscopic research plays a key role in the identification and analysis of interstellar ices and their structure. To date, a number of molecules have been positively
   identified in interstellar ices, either as pure, mixed or layered ice structures.}
   {Previous laboratory studies on H$_{2}$O:CO ices have employed a \textquoteleft mix and match' principle and describe qualitatively how absorption bands behave for different physical conditions. The aim of this study is to quantitatively characterize the absorption bands of solid CO and H$_{2}$O, both pure and in their binary mixtures, as a function of partner concentration
   and temperature.}
{Laboratory measurements based on Fourier transform infrared
transmission spectroscopy are performed on binary mixtures of H$_2$O and CO ranging from 1:4 to 4:1.}
{A quantitative analysis of the band profiles and band strengths of \water in CO ice, and
vice versa, is presented and interpreted in terms of two models. The
results show that a mutual interaction takes place between the two
species in the solid, which alters the band positions and band strengths. It is found that the band strengths of the \water bulk stretch, bending and libration vibrational bands decrease linearly by a factor of up to 2 when the CO concentration is increased from 0 to 80\%. By contrast, the band strength of the free OH stretch increases linearly. The results are compared to a recently performed quantitative study on H$_{2}$O:\co ice mixtures. It is shown that for mixing ratios of 1:0.5 H$_{2}$O:X and higher, the \water bending mode offers a good tracer to distinguish between \co or CO in \water ice. Additionally, it is found that the band strength of the CO fundamental remains constant when the water concentration is increased in the ice. The integrated absorbance of the 2152 cm$^{-1}$ CO feature, with respect to the total integrated CO absorption feature, is found to be a good indicator of the degree of mixing of CO in the H$_2$O:CO laboratory ice system. From the change in the \water absorption band strength in laboratory ices upon mixing we conclude that astronomical water ice column densities on various lines of sight can be underestimated by up to 25\% if significant amounts of CO and \co are mixed in.} 
{}


\maketitle
%

\section{Introduction}

Water and carbon monoxide are common constituents in vast regions of
space, both in the gas phase and in ices. Interstellar water ice was
first identified in 1973 via a strong band at 3.05 $\mu$m and
unambiguously assigned to water ice following comprehensive
laboratory work \citep{merrill76,leger79,Hagen79}. Meanwhile, it has become
clear that \water ice is the most abundant ice in space. The OH stretching mode at 3.05 $\mu$m and the \water bending mode at 6.0 $\mu$m are detected in many lines of sight \citep[e.g. ][]{Willner82,Tanaka90,Murakawa00,Boogert00,keane01,Gibb04,Knez05} and in many different environments, ranging from quiescent dark clouds to dense star forming regions and protoplanetary disks \citep{whittet88,tanaka94}. It has been a long-standing problem that the intensity ratio of these two water bands in astrophysical observations is substantially different from values derived from laboratory spectra of pure \water ice. In recent years it has been proposed that this discrepancy may be due to contributions of other species, in particular more complex organic ices, to the overall intensity of the 6 $\mu$m band \citep{Gibb02}. An alternative explanation is that the band strengths change due to interaction of \water molecules with other constituents in the ice. In both high-mass and low-mass star forming regions, CO is - together with \co - the most dominant species that could mix with H$_{2}$O. In a recent study on H$_{2}$O:CO$_{2}$ ices, \citet{oberg06} showed indeed significant band strength differences between pure and mixed \water ices. The present study extends this work to CO containing water ice.

CO accretes onto dust grains around 20 K \citep{sandford88,acharyya07} and plays a key role in solid state astrochemical processes, e.g., as a starting point in hydrogenation reactions that result in the formation of formaldehyde and methanol \citep{watanabe02,hiraoka02,watanabe04,fuchs07}. A strong absorption centered around 2139 cm$^{-1}$ was assigned to solid CO by \cite{soifer79}, again following thorough laboratory infrared work. Further efforts in the laboratory have shown that CO molecules can be intimately mixed, either with molecules that possess
the ability to form hydrogen bonds, such as H$_{2}$O, NH${_3}$ and
CH${_3}$OH - often referred to as \textquotedblleft polar" ices - or with molecules
that can only participate in a van der Waals type of bond, such as
CO itself, \co and possibly N${_2}$ and O${_2}$ - so-called \textquotedblleft non-polar" ices. In laboratory mixtures with \water and CO, the two forms are distinguished spectroscopically; the double Gaussian peak structure for the CO stretch fundamental can be decomposed in Gaussian profiles at 4.647 $\mu$m (2152 cm$^{-1}$) and 4.675 $\mu$m (2139 cm$^{-1}$), attributed to the polar and non-polar component, respectively \citep{sandford88,jenniskens95}. On the contrary, pure CO measured in the laboratory exhibits a single Lorentzian band, which is located around 2139 cm$^{-1}$. This Lorentzian absorption profile can be further decomposed into three Lorentzian components centered around 2138.7, 2139.7 and 2141.5 cm$^{-1}$ \citep{fraser07}. 

In astronomical spectra, the 2139 \wav feature has been considered as an indicator of CO in \water poor ice, and the 2136 \wav feature as CO in \water rich environments \citep{Tielens91}. More recently it was found that the astronomical CO profiles can be decomposed into three components at 2136.5 cm$^{-1}$, 2139.9 cm$^{-1}$ and 2143.7 cm$^{-1}$, with the 2139.9 \wav feature ascribed to pure CO ice, and the 2143.7 \wav feature ascribed to the longitudinal optical (LO) component of the vibrational transition in pure crystalline CO \citep{pontoppidan03}. \citet{Boogert02} proposed that the astronomically observed peak at 2143 \wav can originate from CO:\co mixtures, but this identification is still controversial \citep{Broekhuizen06}. The assignment of the 2136.5 \wav feature in these phenomenological fits remains unclear. It should be noted that laboratory and astronomical data differ slightly in peak position, largely due to the fact that grain shape effects play a role for abundant ice molecules like CO and H$_2$O.

Recently, elaborate laboratory work and \itshape ab initio \upshape calculations on mixtures of CO and \water have shown that the absorption around 2152 cm$^{-1}$ results from CO being bound to the dangling OH site in \water ice \citep{halabi04}. Surprisingly enough, this absorption has never been observed in the interstellar medium \citep[e.g. ][]{pontoppidan03}. The non-detection of this feature has been explained by other molecules blocking the dangling OH site, which is therefore unavailable to CO. An extension of this explanation is that the binding sites are originally populated by CO, but that this has been processed to other molecules, such as \co or methanol \citep{fraser04}. Furthermore, it has been shown that the number of dangling OH sites decreases upon ion irradiation, which in turn results in a reduction of the integrated intensity of the 2152 \wav feature \cite[ and references therein]{palumbo06}. The 2136-2139 cm$^{-1}$ feature is ascribed to CO bound to fully hydrogen bonded water molecules \citep{halabi04}.
 

Since CO and \water are among the most abundant molecules in the interstellar medium, mixed CO and \water ices have been subject to many experimental and theoretical studies \citep[e.g. ][]{Jiang74,Hagen81,hagen83,halabi04,fraser05}. For example, the behavior of the 2136-2139 \wav CO stretching band has been quantitatively studied as a function of temperature and its band width and position have been studied as a function of \water concentration in binary mixtures, but containing only up to 25\% of CO \citep{schmitt89b,Schmitt89}. Furthermore, water clusters have been studied in a matrix of CO molecules with a ratio of 1:200 H$_{2}$O:CO. This has resulted in a tentative assignment of \water monomers and dimers and the conclusion that \water forms a bifurcated dimer structure in CO \citep{Hagen81}. Other studies have focussed on Temperature Programmed Desorption (TPD) combined with Reflection Absorption Infrared Spectroscopy (RAIRS) of mixed and layered CO/\water systems, enhancing greatly our knowledge on their structures and phase transitions \citep{Collings03a, Collings03b}. Nevertheless, a full quantitative and systematic study on the behavior of \water in CO ice, and vice versa, with straight applications to astronomical spectra, is lacking in the literature. This is the topic of the present work.

The desorption temperatures of CO and \water differ by as much as 145 K under laboratory conditions. However, H$_{2}$O/CO ices are expected to play a role in astronomical environments at temperatures not only well below the desorption temperature of CO at 20 K \citep{fuchs07}, but also well above the desorption temperature of pure CO ice, since CO can be trapped in the pores of H$_{2}$O ice \citep{Collings03a}. Thus far, both species have been observed together in lines of sight. It is often concluded from the non-detection of the 2152 \wav feature that \water and CO are not intimately mixed in interstellar ices. On the other hand, in some lines of sight CO is trapped in pores of a host matrix, as evidenced by the detection of the 2136 \wav CO feature \citep{pontoppidan03}. It is plausible that this trapping results from heating of a mixture of CO and a host molecule. Accordingly, we have also performed some experiments as a function of temperature.

In this work, the effect of CO on the \water vibrational fundamentals is compared to the effect of \co on these modes, as studied recently by \cite{oberg06}. A comparison between the \water bending mode characteristics in CO and \co containing ices illustrates the sensitivity of this mode to the molecular environment. In addition, this work provides a unique laboratory tool for investigating the amount of CO mixed with water.

The outline of this manuscript is as follows. In Sect. 2 the
experimental setup is described and the data analysis is explained.
Section 3 is dedicated to the influence of CO on the water
vibrational modes, as well as the influence of water on the CO bands. In Sect. 4, the astrophysical relevance is discussed and the conclusions
are summarized in Sect. 5.


\section{Experiment and data analysis}

The experimental setup used for the measurements has been described
in detail in \cite{gerakines95}. It consists of a high vacuum setup ($\approx 10^{-7}$ Torr) in which ices are grown on a CsI window at a temperature of 15 K. The window is cooled down by a closed cycle He refrigerator and the sample temperature is controlled by resistive heating. A Fourier Transform InfraRed (FTIR) spectrometer is used to record ice spectra in
transmission mode from 4000 to 400 cm$^{-1}$ (2.5-25 $\mu$m) with a resolution of 1 cm$^{-1}$.

The sample gas mixtures are
prepared in glass bulbs, using a glass vacuum manifold. The
bulbs are filled to a total pressure of 10 mbar, which is well below
the water vapor pressure. The base pressure of the manifold is
better than 10$^{-4}$ mbar, resulting in negligible contamination
levels. A sample of CO (Praxair 99,999\%) is used without further
purification. Deionized water, further purified by three freeze pump
thaw cycles, is used for the H$_2$O:CO mixtures. Mixtures with
different ratios H$_2$O:CO are prepared in the vacuum manifold and
 the resulting depositions are listed in Table \ref{table1}. The growth rate onto the ice is determined by setting the
exposure to $\sim$10$^{16}$ molecules cm$^{-2}$ s$^{-1}$. Assuming a monolayer surface
coverage of 10$^{15}$ molecules$\cdot$cm$^{-2}$ and a sticking probability
of 1, this results in a growth rate of 10 L$\cdot$s$^{-1}$ (L = Langmuir). In the experiments where the effect of CO on the water ice vibrational modes is investigated, the water exposure has been kept constant with about 3000 L of water ice for the different mixtures to facilitate a one-on-one comparison between all samples. In the experiments where the effect of the \water on the CO modes has been investigated, the total amount of deposited CO is kept constant at 3000 L (Table \ref{table1}).

Three independent measurements are performed for 1:1 H$_2$O:CO
mixtures. These measurements allow for an estimate of the error in
the experiment due to mixing of the gas, deposition of the sample
and other errors that may occur. A conservative
error of $\approx$10\% on the mixing ratios is deduced from these experiments.
Additionally, two test measurements are performed for samples of 1000
and 10000 L, to check for layer thickness dependencies
(Table \ref{table1}).

The infrared transmission spectra are taken using a Biorad FTS40 spectrometer. A total of 256 spectra are
acquired and averaged for
each sample measurement. Background measurements are performed prior
to each of the measurements and are subtracted to reduce the noise
level. The spectra are further processed using IDL (Interactive Data Language) in order to flatten the baseline. This is done by fitting a second order
polynomial through a set of five points, which are visually chosen
well away from absorption features. The data reduction does not lead
to a distortion of the absorption profiles.

\begin{table}
\begin{minipage}[]{\columnwidth}
\caption{Ice mixtures and resulting deposition thicknesses used in this work. Column A denotes the molecule of which the deposited amount is kept constant, and column B indicates the molecule that is mixed in. The first series is used for determining the effect of CO on the \water band strengths and profiles. The second series is used to determine the effects of \water on the CO band strengths and profiles.} \label{table1} \centering
\renewcommand{\footnoterule}{}
\begin{tabular}{l c c c}
\hline\hline
 Composition&A (L\footnote{1 L (Langmuir) = 1 $\times$ 10$^{-6}$
 Torr s $\approx$ 1 monolayer of molecules})&B (L)&Total (L)\\
\hline
pure H$_2$O&3000&0&3000\\
pure CO&0&3000&3000\\
\hline
H$_2$O:CO 1:0.25&3000&750&3750\\
H$_2$O:CO 1:0.5&3000&1500&4500\\
H$_2$O:CO 1:1&3000&3000&6000\\
H$_2$O:CO 1:2&3000&6000&9000\\
H$_2$O:CO 1:4&3000&12000&15000\\
H$_2$O:CO 1:1&10000&10000&20000\\
H$_2$O:CO 1:1&1000&1000&2000\\
\hline
CO:H$_2$O 1:0.25&3000&750&3750\\
CO:H$_2$O 1:0.5&3000&1500&4500\\
CO:H$_2$O 1:1&3000&3000&6000\\
CO:H$_2$O 1:2&3000&6000&9000\\
CO:H$_2$O 1:4&3000&12000&15000\\
CO:H$_2$O 1:1&10000&10000&20000\\
CO:H$_2$O 1:1&1000&1000&2000\\
\hline
\end{tabular}
\end{minipage}
\end{table}

\begin{figure}
\includegraphics[width=9cm]{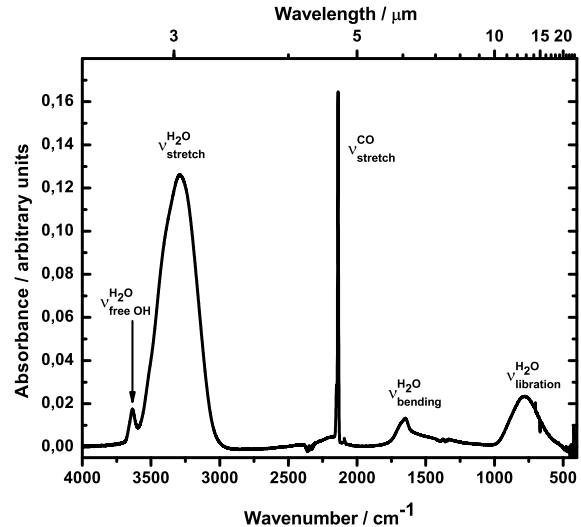} \caption{A typical baseline and
background corrected ice spectrum for a H$_2$O:CO = 1:1 mixture. The
measurement is performed at 15 K. The vibrational modes in the H$_{2}$O:CO ice are indicated.}
\label{fig1}
\end{figure}

The absorption band strengths for the three modes of pure water ice at
15 K are well known from literature \citep{gerakines95}. The adopted
values are 2$\times$10$^{-16}$, 1.2$\times$10$^{-17}$ and
3.1$\times$10$^{-17}$ cm molecule$^{-1}$ for the stretching ($\nu\rm_{stretch}$=3279
cm$^{-1}$ or 3.05 $\mu$m), bending ($\nu\rm_{bend}$=1655 cm$^{-1}$ or 6.04 $\mu$m) and libration mode ($\nu\rm_{lib}$=780
cm$^{-1}$ or 12.8 $\mu$m), respectively (see Fig.\ref{fig1}). Calculating the
integrated absorption bands over the intervals listed in Table \ref{table2} for the mixtures and scaling them to the
integrated band strength for pure water ice, allows for a deduction of
the band strengths for the water ice bands in the mixture via:

 \begin{equation}
A_{\rm H{_2}O:CO{=}1:x}^{\rm band}{=}{{\int_{\rm band}{I_{\rm H{_2}O:CO{=}1:x}}}{\times} \frac{ A_{\rm H{_2}O}^{\rm band}}{\int_{\rm band}{I_{\rm H{_2}O}}}}
 \label{eqn1}
 \end{equation}

\noindent where $A_{\rm H{_2}O:CO{=}1:x}^{\rm band}$ is the
calculated band strength for the vibrational water mode in the 1:x
mixture, ${I_{\rm H{_2}O:CO{=}1:x}}$ its integrated area, ${ A_{\rm
H{_2}O}^{\rm band}}$ the band strengths for the water
modes as available from literature and ${\int_{\rm band}{I_{\rm H{_2}O}}}$ the integrated area
under the vibrational mode for the pure water sample. The free OH
stretching mode, \textquoteleft the fourth band\textquoteright, is scaled to the stretching mode for pure water
since this absorption is absent in the spectrum of pure H$_2$O ice.

Integration limits used throughout the experiment are listed in
Table \ref{table2}. Integrated areas relative to the integrated area
of the pure water stretching mode, $ A/A$ \begin{tiny}pure H$_{2}$O
stretch\end{tiny}, are investigated as a function of CO
concentration. For the sample with the most mixed in CO, i.e., the 1:4 H$_{2}$O:CO mixture, an analysis in terms of cluster formation
is given. In addition, the influence of temperature on the water stretching mode is studied. The measured spectra for the H$_2$O:CO mixtures are available online at the Leiden ice database: http://www.strw.leidenuniv.nl/$\sim$lab/databases.

\begin{table}
\begin{minipage}[t]{\columnwidth}
\caption{The measured peak positions and the integration bounds in
cm$^{-1}$ used to compute the integrated intensities of the
H$_2$O bands. The values between brackets indicate the $\mu$m values.}
\label{table2} \centering
\renewcommand{\footnoterule}{}
\begin{tabular}{l c c c c}
\hline\hline
 &&&\multicolumn{2}{c}{Integration bounds}\\ 
Species&Assignment&Peak&Lower&Upper\\
\hline
H$_{2}$O&$\nu_{\rm libration}$&780 (12.8)&500 (20.0)&1100 (9.09)\\
&$\nu_{\rm bend}$&1655 (6.04)&1100 (9.09)&1900 (5.26)\\
&$\nu_{\rm stretch}$&3279 (3.05)&3000 (3.33)&3600 (2.78)\\
&$\nu_{\rm free \ OH}$&3655 (2.73)&3600 (2.78)&3730 (2.68)\\
\hline
CO \footnote{The integrations for the two CO bands are performed using Gaussian fits.} &$\nu_{\rm 'stretch'}$&2139 (4.68)&2120 (4.72)&2170 (4.61)\\
&$\nu_{\rm 'polar'}$&2152 (4.65)&2120 (4.72)&2170 (4.61)\\
\hline
\end{tabular}
\end{minipage}
\end{table}

\section{Results}

\subsection{Influence of CO on water bands}

\begin{figure*}
\centering
\includegraphics[width=18cm]{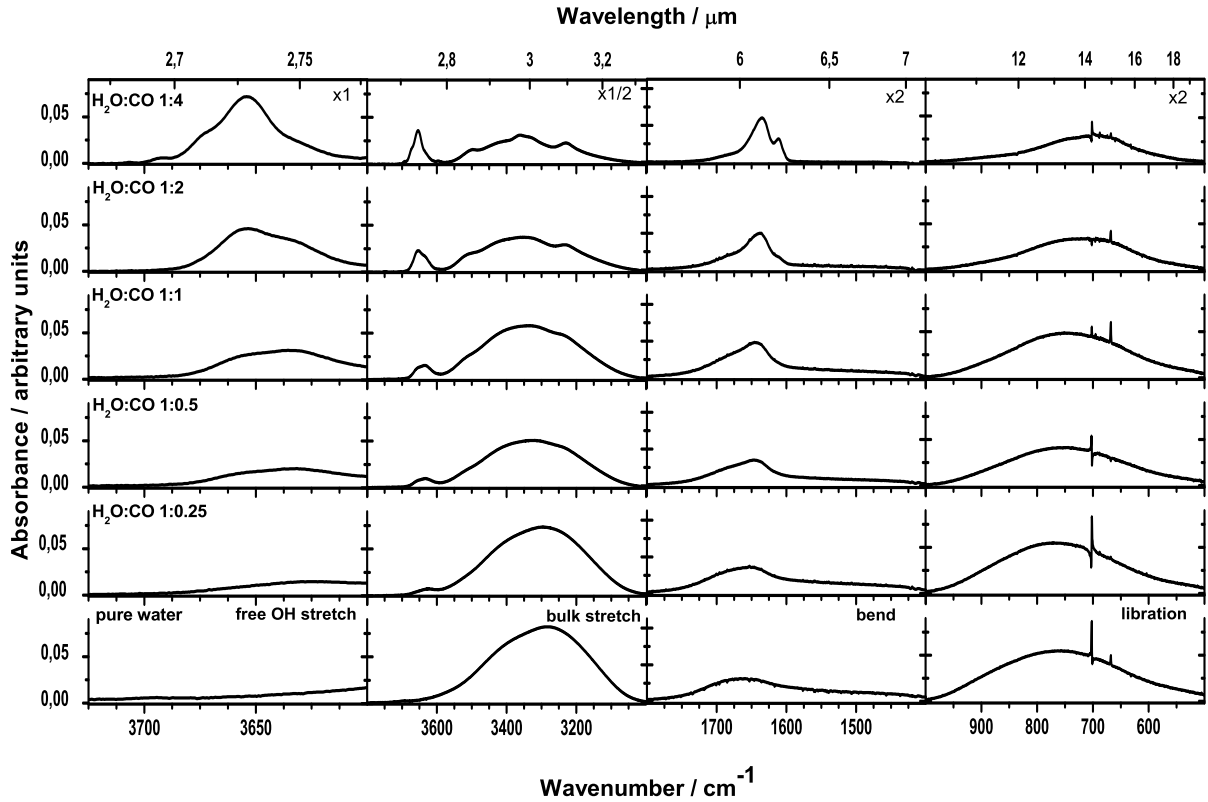}
\caption{Combined spectra of the four modes for water ice for the six
measured compositions (see Table \ref{table1}), ranging from pure water ice (bottom) to a 1:4
H$_2$O:CO mixture (top figures). The spectra are taken at a
temperature of 15 K. Note that the wavelength ranges for separate modes are different. The small structures on the libration mode are experimental artifacts. }
 \label{fig3}
\end{figure*}

\begin{figure}
\centering
\includegraphics[width=9cm]{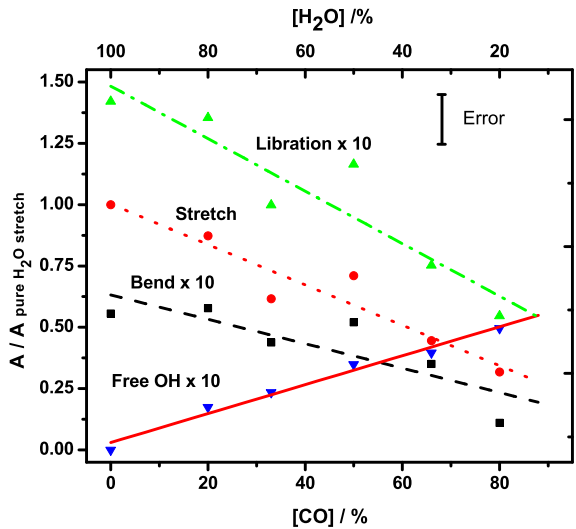} 
\caption{The integrated intensity
of the water vibration modes relative to the integrated intensity of
the pure water stretch mode plotted versus the concentration of CO
in the sample ice. It should be noted that the plots for the bending, free OH and libration mode have been multiplied by a factor of ten to facilitate the display and that the stretch mode is the one that is actually most dependent on the concentration. The four water modes show to first order a linear dependence on the CO
concentration. The estimated
error in the data amounts to 10\%.} \label{fig2}
\end{figure}

In Fig. \ref{fig3}, the four \water ice fundamentals are shown for different compositions. Similar to CO$_2$ \citep{oberg06}, CO has a clear influence on the
water ice absorption bands compared to the pure \water ice. This effect is different for each of the
four bands. The bulk stretch mode is
most strongly affected; the band strength for this mode decreases by
more than a factor of 2 when the CO fraction is raised from 0\% to 80\%. The
band strength of the free OH stretch, which is absent when no CO is
mixed in, is greatly enhanced with concentration. The libration mode gradually looses intensity when the amount of CO
in the ice is increased and the peak of the absorption band shifts to lower energy. The integrated areas of the four water
modes are scaled to the pure water stretching mode and plotted
versus the CO concentration in Fig. \ref{fig2}. A linear function $A_{\rm eff} = a\cdotp{\rm[CO]} + b$ is fitted through the data points of the four water modes. The \itshape a \upshape coefficient indicates the strength of increase/decrease of the band strength, and the \itshape b \upshape coefficient indicates the band strength of water relative to the pure stretching mode when no impurities are mixed in. There exists some deviation between the data points and the fit function, which is most probably due to the deposition accuracy, but this deviation is within the experimental error of 10\%. A clear trend in all four modes is observed. In Table \ref{table3}, the linear fit coefficients are listed for the H$_2$O:CO binary mixtures. The linear coefficient for the \water stretching mode is highest and negative, indicating the strongest decrease in band strength. The free OH stretching mode has a positive linear coefficient indicating that this is the only mode to increase in intensity upon CO increase. A comparison with recently obtained data for H$_{2}$O:\co ices shows the same trend. Apart from the bending mode, all effects are more pronounced in the \co mixtures, i.e., the absolute values of the \itshape a \upshape coefficient are larger by a factor of 1.3 - 2. This is related to the actual interactions in the ice and work is in progress to study such effects in more detail for a large number of species from a chemical physics perspective.

\begin{figure}
\centering
\includegraphics[width=9.35cm]{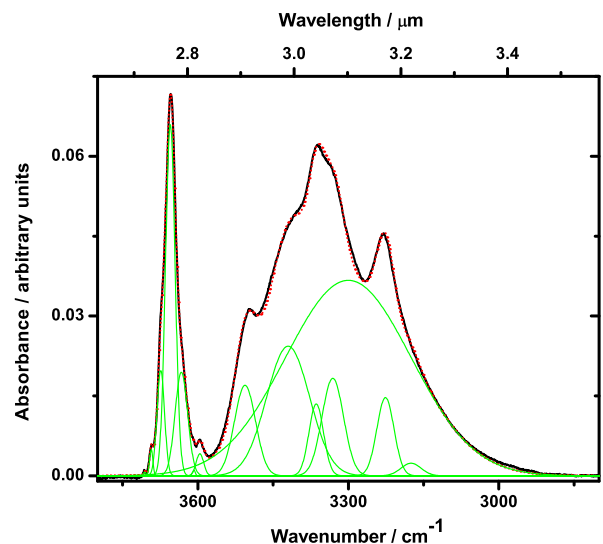} \caption{Measured (indicated in black) and fitted spectrum (indicated in red) of a 1:4 H$_2$O:CO mixture in the frequency range of the \water bulk stretch and free OH stretch modes at a temperature of 15K. The water stretching mode clearly shows substructure. An excellent fit is obtained by a superposition of the bulk stretch mode and Gaussian functions representing smaller \water clusters (indicated in green) in the matrix material \citep{ohno05}. The fitted peak positions and the positions from \citet{ohno05} are listed in Table \ref{table4}. Please refer to the online version for the color coding.} \label{fig4}
\end{figure}

The free OH mode, the water stretching mode and the water bending mode start showing substructure superposed onto the bulk absorption profile upon increase of the fraction of CO in the ice mixture (Fig. \ref{fig4}). The absorptions of the bulk modes are still clearly apparent beneath the substructure. For the stretching mode this absorption shifts from 3279 \wav to a higher wavenumber of 3300 cm$^{-1}$. The peak absorption for the bulk bending mode shifts from 1655 \wav for pure \water to 1635 \wav for the 1:4 H$_2$O:CO mixture. The libration mode is also red-shifted upon CO concentration increase. For the pure \water ice spectrum this mode is located at 780 cm$^{-1}$, while for the highest partner concentration it appears at 705 cm$^{-1}$. The free OH stretch gradually increases in frequency. The peak absorption shifts from 3636 \wav for the 1:0.25 to 3655 \wav for the 1:4 H$_2$O:CO mixture, which corresponds to a blue shift of 19 cm$^{-1}$. 

\begin{table*}
\caption{Resulting linear fit coefficients for the H$_2$O:CO
mixtures. The coefficients indicate the strength of the interaction
between CO and the \water host molecules in the matrix for mixtures that are
deposited at a temperature of 15 K. The corresponding values for H$_2$O:\co ice mixtures are listed for a comparison.}\label{table3} 
\centering
\begin{tabular}{l cccc}
\hline\hline
 &&\multicolumn{2}{c}{Linear Coefficients}\\
 mixture&\water mode&constant (\itshape b \upshape)&slope (\itshape a \upshape)&$R{^2}$\\
 &&[10$^{-16}$ cm molecule$^{-1}$]&[10$^{-19}$ cm molecule$^{-1}$]&\\
\hline
H$_2$O:CO&$\nu_{\rm libration}$&0.30$\pm$0.02&$-$2.1$\pm$0.4&0.93\\
&$\nu_{\rm bend}$&0.13$\pm$0.02&$-$1.0$\pm$0.3&0.84\\
&$\nu_{\rm stretch}$&2.0$\pm$0.1&$-$16$\pm$3&0.95\\
&$\nu_{\rm free \ OH}$&0&1.2$\pm$0.1&0.99\\
\hline
H$_2$O:\co $^{a}$&$\nu_{\rm libration}$&0.32$\pm$0.02&$-$3.2$\pm$0.4&0.99\\
&$\nu_{\rm bend}$&0.14$\pm$0.01&$-$0.5$\pm$0.2&0.81\\
&$\nu_{\rm stretch}$&2.1$\pm$0.1&$-$22$\pm$2&0.99\\
&$\nu_{\rm free \ OH}$&0&1.62$\pm$0.07&0.99\\
\hline
$^{a}$ From \cite{oberg06}
\end{tabular}
\end{table*}

The substructure, which is superimposed on the bulk stretching mode around 3300 cm$^{-1}$, has been previously assigned to (H$_2$O)$_n$ water clusters in the ice. These assignments are based on matrix spectroscopic data of \water in a matrix of N$_{2}$ + O$_{2}$ (75:25) \citep{ohno05}. Comparable data for \water in a matrix of CO are not available. Nevertheless, in Fig. \ref{fig4} it is shown that an excellent fit is obtained when peak
positions from matrix spectroscopic data are used, assuming that the peak
positions are shifted, which is indicative for the difference in interaction between \water and CO compared to \water and N$_{2}$ + O$_{2}$. For each of the contributions, the peak position, bandwidths (Full-Width-at-Half-Maximum, FWHM) and integrated area, are summarized in Table \ref{table4} and compared to previous results obtained by \cite{ohno05}. Some absorptions are red-shifted and other absorptions are blue-shifted compared to absorptions of \water clusters in a N$_{2}$ + O$_{2}$ matrix. Note that the relative \water concentrations in the present work are substantially higher than in the matrix experiments by \cite{ohno05}. As a consequence, larger \water clusters are more pronounced in our spectra. 

\begin{figure}[t]
\centering
\includegraphics[width=9.1cm]{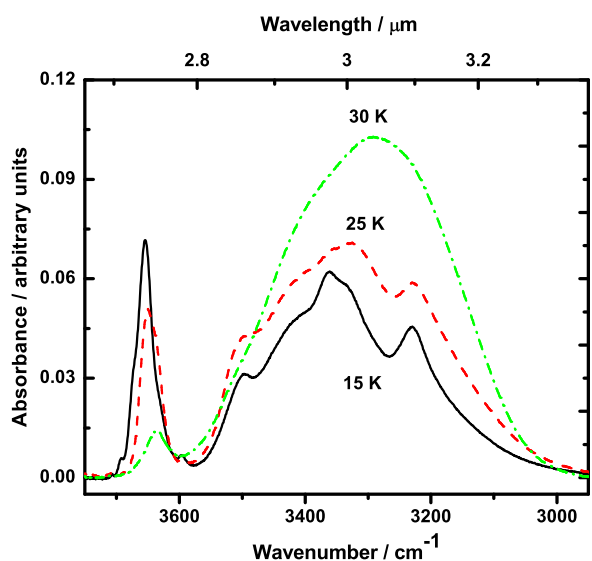} \caption{Temperature dependence
of larger clusters of water molecules in a matrix of H$_2$O:CO =
1:4. For increasing temperature the substructure gives way to the
 bulk stretch mode when CO evaporates.} \label{fig5}
\end{figure}

Additionally, the temperature of the sample plays a role on the band strengths. A clear effect is
encountered when the temperature is slightly increased, 
i.e., from 15 to 25 K, close to the desorption temperature of CO. At this temperature, the CO molecules in
the matrix start gaining enough energy to become mobile. The
mobility in the matrix allows water clusters to find partners for hydrogen
bindings and to reorganize themselves to form a stronger bulk
hydrogen bonded network, as indicated by the increased bulk stretch
mode band strength and decreasing intensity of the substructure. Figure \ref{fig5} nicely shows the transition from \water clusters embedded in a matrix of CO to the formation of a bulk hydrogen bonded network.

\begin{table*}
\caption{Line positions, FWHMs and integrated areas of the Gaussian functions
fitted to the 1:4 H$_2$O:CO water bulk stretch and free OH stretch spectrum. The assignment is based on
both Density Functional Theory (DFT) calculations and experimental values from \cite{ohno05} obtained in a N$_{2}$/O$_{2}$ matrix.} \label{table4} \centering
\begin{tabular}{l ccccc}
\hline\hline
 Mode&Position$^{a}$&Position$^{b}$&$\Delta$ Position$^{b}$&FWHM$^{b}$&Area$^{b}$\\
 &(cm$^{-1}$)&(cm$^{-1}$)&(cm$^{-1}$)&(cm$^{-1}$)&(a.u. $\cdot$ cm$^{-1}$)\\
\hline Free OH (ring) cyclic-pentamer, cyclic-trimer free OH, clusters hexamer&3688&3674&15&15&0.371\\
Asymmetric stretch H-acceptor dimer&3715&3693&22&7&0.040\\
Asymmetric stretch monomer&3715&3706&9&5&0.043\\
Symmetric stretch monomer&3635&3608&27&5&0.074\\
Symmetric stretch H-acceptor dimer&3629&3596&33&15&0.078\\
Bulk free OH stretch&-----&3655&-----&19&1.57\\
Bulk free OH stretch&-----&3633&-----&25&0.609\\
\hline
H-bonded OH stretch "chair" hexamer &3330&3331&{-}1&43&0.987\\
H-bonded OH stretch "cage" hexamer  &3224&3226&{-}2&34&0.626\\
H-bonded OH stretch "prism" hexamer &3161&3175&{-}14&40&0.119\\
H-bonded OH stretch pentamer &3368&3364&4&29&0.491\\
H-bonded OH stretch "book1, cage" hexamer &3450&3420&30&86&2.62\\
Bulk stretching mode&-----&3300&-----&250&11.5\\
H-bonded OH stretch ring cyclic trimer&3507&3506&1&42&0.894\\
\hline
$^{a}$ \cite{ohno05}, $^{b}$ This work.
\end{tabular}

\end{table*}

\begin{figure}[t]
\centering
\includegraphics[width=10cm]{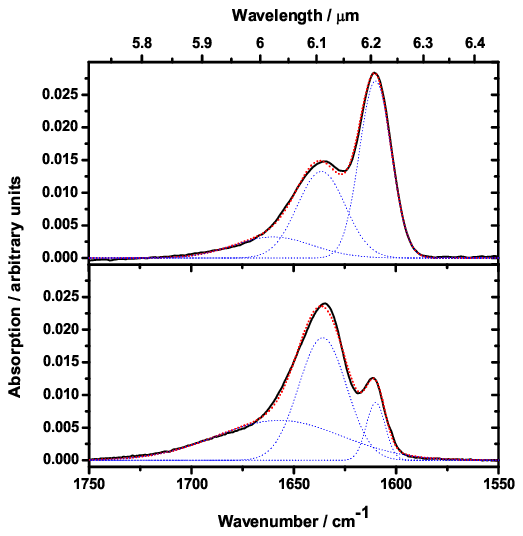} \caption{The spectral differences in the water bending mode profile for a 1:4 H$_2$O:\co ice mixture (top) \citep{oberg06} and a 1:4 H$_2$O:CO mixture (bottom). The black bold (overall) spectra indicate the laboratory spectra and coincide with the fitted spectra (red dotted) consisting of the three Gaussian curves, indicated by the blue dotted lines. (See online version)} \label{fig6}
\end{figure}

Figure \ref{fig6} shows the different effect of CO and \co on the \water bending mode in ice for the 1:4 H$_{2}$O:\co and 1:4 H$_{2}$O:CO mixtures. The intensity ratio of the main peaks is actually reversed in the two mixtures. The differences between the two mixtures start showing up from a mixing ratio of 1:0.5 H$_{2}$O:X and become more pronounced for higher CO and \co concentrations. In addition, the CO mixtures exhibit a stronger broad underlying feature, which is visualized by the Gaussian fit in Fig. \ref{fig6}. In other words, a detailed study of the \water bending and stretching modes may provide additional information on whether CO or \co dominates in the ice. The free OH stretching mode is also affected differently by the two molecules. In the H$_{2}$O:\co mixtures, the mode is more shifted to higher wavenumbers. For the 1:1 H$_{2}$O:\co mixture the peak position is 3661 cm$^{-1}$, compared to 3635 \wav for the H$_{2}$O:CO mixtures.

Thicker and thinner layers of the mixtures have been measured to
check for thickness dependence. We conclude that within our experimental error limit, ice thickness does not play a significant role in the behavior of the relative band
strengths. This conclusion is supported by the observation that identical mixing
ratios in the two measurement series (H$_2$O/CO and CO/H$_2$O) show the same (scaled) spectroscopic behavior for different total ice thicknesses (Table \ref{table1}).

\subsection{Influence on the CO band}

\begin{figure}[t]
\centering
\includegraphics[width=9cm]{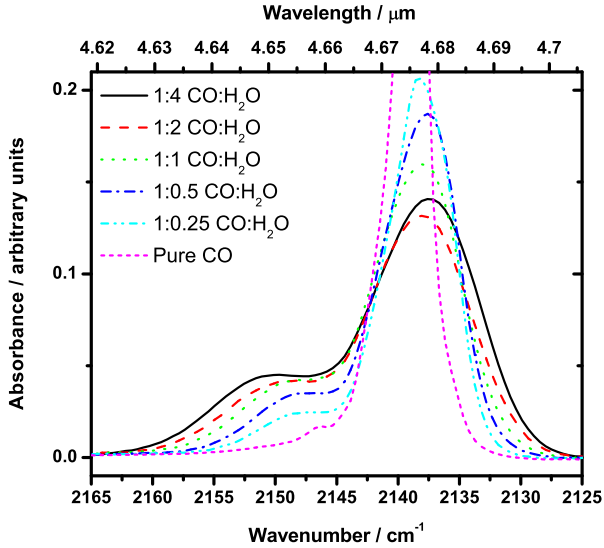} \caption{Illustration of the behavior of the CO stretch fundamental upon increase of the concentration of H$_{2}$O in CO. The total integrated band strength remains unchanged within the experimental error, although the maximum intensity of the absorption decreases strongly. The y-axis is cut off for the pure CO mode to make a clearer distinction between \textquoteleft non-polar' and \textquoteleft polar' components of the CO absorption for the CO:H$_{2}$O mixtures.} \label{fig7}
\end{figure}

\begin{figure}[t]
\centering
\includegraphics[width=9.2cm]{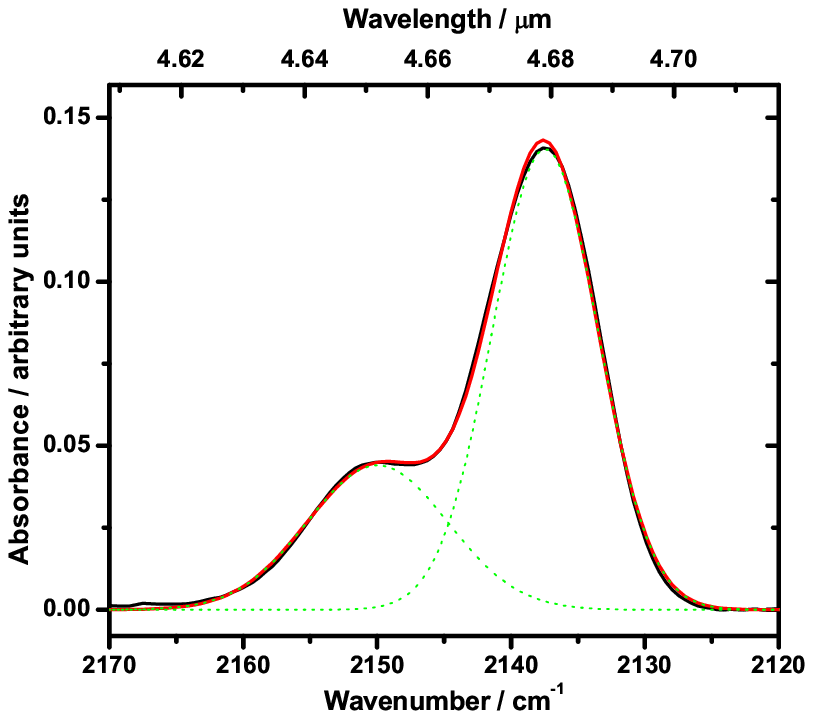} \caption{Gaussian fit of the
CO stretch mode in a CO:H$_{2}$O = 1:0.25 mixture. The black bold spectrum is the
measured laboratory spectrum that is reproduced (red line) by adding the two Gaussian components centered
around 2138.2 cm$^{-1}$ and 2147.5 cm$^{-1}$ (green dotted lines). These are attributed to CO in a
\textquoteleft non-polar' and \textquoteleft polar' environment. (See online version)} \label{fig8}
\end{figure}

Section 3.1 shows that mixing CO into a water mixture affects the band
strengths of the water vibrational modes. Reciprocally, the CO stretch
mode is also altered when water is added to CO ice. This is seen in the experiments where the amount of deposited CO is kept constant (Table \ref{table1}). When water is mixed into the CO ice, the absorption changes from a Lorentzian profile to a Gaussian profile. Furthermore, the second CO absorption at 2152 cm$^{-1}$, ascribed to CO bound to the \water dangling OH sites \citep{fraser04}, manifests itself as a Gaussian profile and increases in integrated intensity upon increase of water concentration. The transition from the pure CO Lorentzian shaped profile to two Gaussian shaped profiles for the mixed CO:\water ices is illustrated in Fig. \ref{fig7}. We have not further decomposed the 2139 \wav component. More information is available from \citet{fraser07}.

Gaussian fits for both CO absorption components are made
for the range of mixtures as listed in Table \ref{table1} at a temperature of 15 K. One typical fit is shown in Fig. {\ref{fig8}. With this example, it is demonstrated that excellent fits are obtained by using the fit parameters as listed in Table \ref{table5}. The behavior of
the integrated area of the 2152 cm$^{-1}$ component compared with the total integrated CO absorption is plotted as a function of CO and \water concentration in Fig. \ref{fig10}. The integrated area for the 2152 \wav component decreases with increasing CO content, and it is undetectable for a pure CO ice. A second order polynomial describes how the polar component behaves with respect to the CO concentration $[x]$ or water concentration $[100-x]$ in the ice over the interval spanning from 20\% CO up to a pure CO ice. Note that the total amount of deposited CO is kept constant. The coefficients of the second order polynomial of the form $y=a\cdot x^2+b\cdot x+c$ \upshape are \itshape a\upshape=$-$0.005, \itshape b\upshape=0.23 and \itshape c\upshape=26.6. For a decreasing amount of water in the sample, the peak position of the 2152 cm$^{-1}$ absorption feature is most strongly affected and decreases gradually to lower wavenumbers, until it reaches 2148 cm$^{-1}$ for the 1:0.25 CO:H$_{2}$O mixture. The FWHM of this band is also affected. It starts at a width of 10.5 cm$^{-1}$ for the 1:4 CO:H$_{2}$O mixture and decreases to 7.5 cm$^{-1}$ for the 1:0.25 CO:H$_{2}$O mixture. The position of the main absorption feature at 2139 cm$^{-1}$ is only slightly affected by increasing the amount of water in the sample and decreases by 1.3 cm$^{-1}$ when going from pure CO ice to the 1:4 CO:H$_{2}$O mixture, i.e., a shift towards the 2136 \wav feature is observed upon dilution (see Fig. \ref{fig9}). Its position is expected to shift even more, to 2136 cm$^{-1}$, for \water concentrations above 80\%. The FWHM of this absorption feature decreases from 8 cm$^{-1}$ for the 1:4 CO:\water to 5 cm$^{-1}$ for the 1:0.25 CO:H$_{2}$O sample. The Lorentzian peak profile of the pure CO absorption exhibits an even smaller FWHM of 2 cm$^{-1}$. An overview of the changes in peak position, FWHM and integrated intensity is given in Table \ref{table5}.

The position of the 2139 \wav absorption feature is also strongly dependent on the temperature of the ice, as illustrated in Fig. \ref{fig9}. The majority of the CO will desorb as the ice is heated above the CO desorption temperature. The remaining CO shows an absorption that is shifted toward 2135 cm$^{-1}$. Thus, a shift from 2139 to 2136
cm$^{-1}$ occurs both by mixing with significant amounts of H$_{2}$O and by heating of CO:H$_{2}$O above 40 K, even for mixtures with modest amounts of H$_{2}$O. From Fig. \ref{fig9}, it becomes clear that the latter effect (i.e. heating) is the more critical one. One should note that the laboratory data presented here cannot be compared one-to-one with the observational data because of shifts caused by grain shape effects.
 
\begin{figure*}[t]
\centering
\includegraphics[width=18cm]{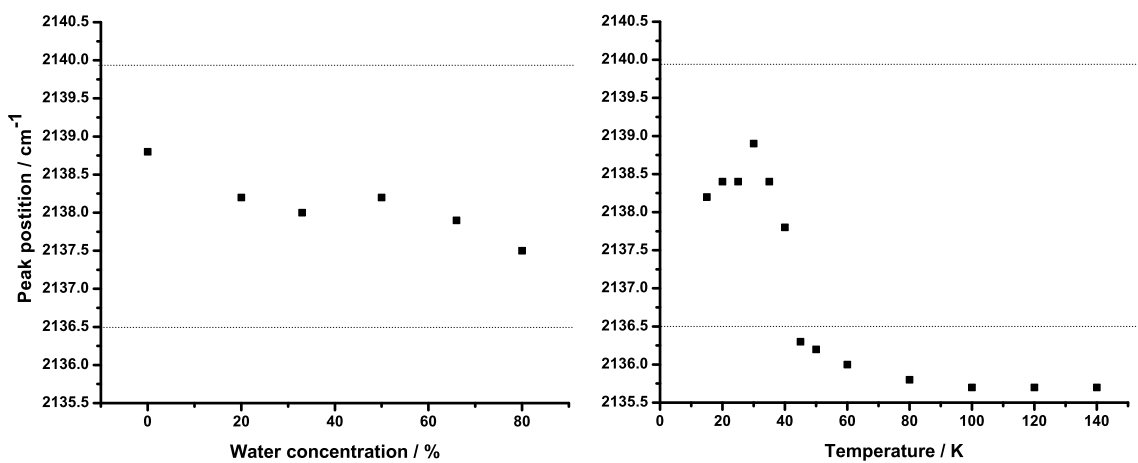} \caption{Change in peak position of the 2139 \wav CO absorption feature as a function of both \water concentration and temperature for the 1:4 H$_{2}$O:CO mixture. The dotted lines indicate the band positions measured for CO in the interstellar medium. } \label{fig9}
\end{figure*}

\begin{table}
\begin{minipage}[]{\columnwidth}
\caption{Lorentzian and Gaussian fit parameters for the CO stretching mode for a constant amount of CO in ice mixtures ranging from 100\% CO to a 1:4 CO:\water mixture.} \label{table5} \centering
\renewcommand{\footnoterule}{}
\begin{tabular}{l c c c c}
\hline 
\hline
 Composition&Position&FWHM&Area\\
 &(cm$^{-1}$)&(cm$^{-1}$)&(a.u. $\cdot$ cm$^{-1}$)\\
\hline 
Pure CO \footnote{Lorentzian profile}&2138.8&2.2&1.9\\
1:0.25 CO:\water&2138.2&5.0&1.31\\
&2147.5&7.5&0.23\\
1:0.5 CO:\water&2138.0&5.8&1.40\\
&2148.1&7.8&0.34\\
1:1 CO:\water&2138.2&6.5&1.28\\
&2148.3&9.0&0.44\\
1:2 CO:\water&2137.9&7.6&1.24\\
&2149.5&9.8&0.49\\
1:4 CO:\water&2137.5&7.9&1.38\\
&2150&10.5&0.58\\
\hline
\end{tabular}
\end{minipage}
\end{table}

\begin{figure}[t]
\centering
\includegraphics[width=9cm]{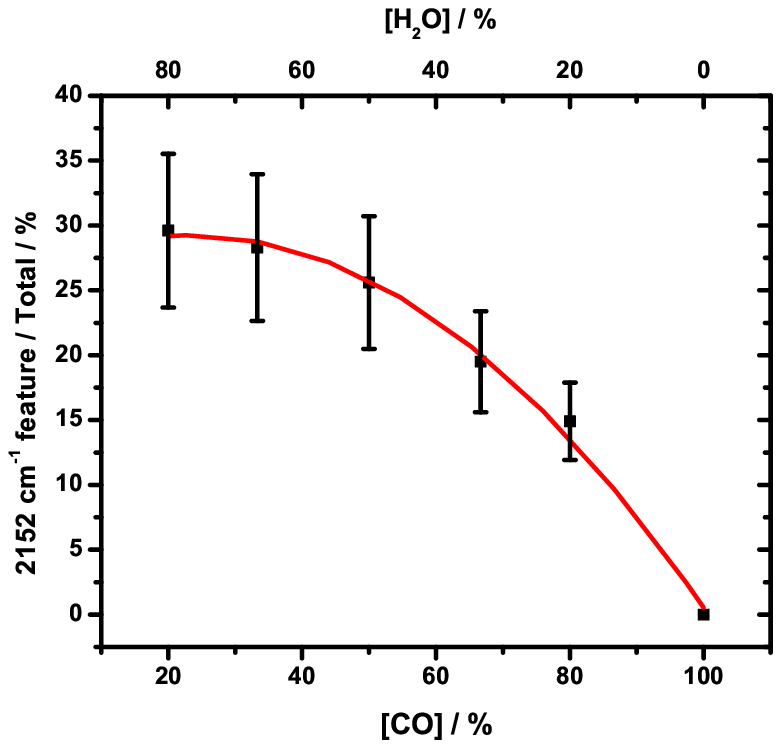} \caption{Absorbance of the 2152 \wav component relative to the total CO absorption as a function of CO
concentration.} \label{fig10}
\end{figure}

\section{Discussion}

CO and \water ice abundances are often derived using the well
known constants from the literature. The experimental work presented
here shows that band strengths deduced from pure ices cannot be used to
derive column densities in interstellar ices without further knowledge on the
environmental conditions in the ice. Concentrations of CO and \co ice as high as 15 and 21\%, respectively, relative to H$_2$O, have been reported towards GL7009S 
\citep{keane01}. If we assume a polar fraction of 75\%, as reported towards W33A, and that the polar fraction of both CO and \co ice are in close contact with the water, the band strength for the \water bending mode is reduced by a factor of $\sim$ 1.25. In other words, the band strength for \water will be smaller and hence the column density of \water will be underestimated if the laboratory data from pure \water ice are adopted from the literature. 

Vice versa, for deriving the column density of CO ice one should also
consider the influence of other molecules in its vicinity. The
integrated absorbance of CO, however, is less strongly influenced by
the presence of H$_{2}$O. The CO absorption decreases in absolute intensity when water is mixed in, but the total integrated band strength is compensated by a broadening of the absorption and the appearance of the 2152 cm$^{-1}$ component, at least in the laboratory spectra. Upon heating, the CO molecules become trapped into pores, as indicated by the 2136 \wav feature. The absorption strength of this band is very sensitive to temperature \citep{schmitt89b}. The CO absorption profile is indicative of the amount of water that is mixed. The percentage of CO mixed into the \water ice is derived in the laboratory from the ratio between the 2152 \wav feature and the total CO absorption (Fig. \ref{fig10}). This allows the derivation of the effective band strength A$_{\rm eff}^{\rm band}$ for \water in a mixture with CO using the linear model proposed in Sect. 3.1, and thus an estimate of the column density for \water via the equation \citep{sandford88}:

 \begin{equation}
N_{\rm corr}{=}\frac{\int{\tau_{\nu}d\nu}}{A_{\rm eff}}
 \label{eqn3}
 \end{equation}

Temperature also affects the
band strength as shown in Fig. \ref{fig5}. The applications of the model presented in Fig. \ref{fig10} are restricted to deduction of column
densities for ices with a temperature of about 15 K, which have not been thermally
processed. It should also be noted, as recently demonstrated by \citet{bisschop07}, that in addition to binary mixtures, tertiary mixtures also have to be taken into account to compare laboratory data with astronomical spectra.

\section{Conclusions}

Based on the experiments described in this manuscript and in recent work on H$_{2}$O:\co mixed ices, we draw the following conclusions regarding the interaction between \water and CO in a solid environment:

\begin{enumerate}
\item The general trend on the band strengths of the four vibrational modes in water ices is similar for H$_{2}$O:CO and H$_{2}$O:\co mixtures with increasing CO or \co concentration. However, quantitative differences exist, reflecting differences in the strength of the interaction, which allow us to distinguish between CO and \co in \water ice, explicitly assuming that the main constituents of the ice are H$_2$O, CO$_2$ and CO.
\item The position of the water free OH stretching mode is particularly indicative of the molecule that is interacting in the matrix (CO vs CO$_{2}$), again under the assumption that we only consider
H$_2$O/CO$_2$/CO ices. The peak position of this mode is 26 \wav red-shifted for a 1:1 H$_{2}$O:CO ice mixture compared to a 1:1 H$_{2}$O:\co ice mixture.
\item In addition, the water bending mode is indicative of the molecule, i.e., CO or CO$_{2}$, that is interacting with the water ice. The relative integrated intensity of the Gaussian components reveals whether \co or CO is mixed into the \water ice. The same restriction mentioned in conclusions 1 and 2 also applies here.
\item Upon increasing the relative amount of CO in the mixture, a clear substructure begins to reveal itself in the bending, free OH stretch and bulk stretching mode. The arising substructure indicates the onset of \water cluster formation in the H$_{2}$O:CO ice. An assignment of the clusters was possible following matrix isolation spectroscopy.
\item The substructure on the stretching mode quickly gives way to the bulk water mode when the temperature is increased close to the desorption temperature of CO. This can easily be depicted by CO molecules becoming mobile and hence allowing single water molecules and larger water clusters to find partners for bulk hydrogen bonding.
\item The ratio 2152 \wav / total integrated CO absorption intensity is a tracer of the amount of CO that is mixed into the laboratory water ice, or vice versa. In astronomical spectra this band has not been observed.
\item \water column densities derived from astronomical spectra can easily be underestimated by as much as 25\% when environmental influences, i.e., CO or \co presence, are not taken into account.
\end{enumerate}

The present systematic study of CO:\water ice, together with recent work on CO$_{2}$:\water ice, provide the tools with which to estimate the mixing ratios of the three most abundant molecules in interstellar ices.  

\begin{acknowledgements}

This work is part of the research programme of the 'Stichting voor
Fundamenteel Onderzoek der Materie (FOM)', which is financially
supported by the 'Nederlandse Organisatie voor Wetenschappelijk
Onderzoek (NWO)'. Additionally, financial support by the 'Greenberg Foundation' and by 'the Netherlands Research School for Astronomy (NOVA)' are gratefully acknowledged.

\end{acknowledgements}


\bibliographystyle{aa}

\end{document}